\documentclass[prc,aps,fonts,amsmath,superscriptaddress,floatfix,twocolumn,nofootinbib]{revtex4-1}
\usepackage{graphicx,float} 
\usepackage{color}
\usepackage{mwe}
\usepackage{braket}
\usepackage{amsfonts}
\usepackage{cancel}
\usepackage{rotating}

\usepackage[caption=false]{subfig}
\usepackage{hyperref}

\usepackage{braket}
\usepackage{array}
\usepackage{ulem}
\newcolumntype{P}[1]{>{\centering\arraybackslash}p{#1}}
\newcommand{\gras}[1]{\boldsymbol{#1}}

\begin{document}

\title{Covariant energy density functionals with and without tensor couplings at the Hartree-Bogoliubov level }

\author{F. Mercier}
\affiliation{IJCLab, Universit\'e Paris-Saclay, IN2P3-CNRS, F-91406 Orsay Cedex, France}

\author{J.-P. Ebran}
\affiliation{CEA,DAM,DIF, F-91297 Arpajon, France}
\affiliation{Universit\'e Paris-Saclay, CEA, Laboratoire Mati\`ere en Conditions Extr\^emes, 91680, Bruy\`eres-le-Ch\^atel, France}

\author{E. Khan}
\affiliation{IJCLab, Universit\'e Paris-Saclay, IN2P3-CNRS, F-91406 Orsay Cedex, France}

\begin{abstract} 
\begin{description}
\item[Background] The study of additional terms in functionals is relevant to better describe nuclear structure phenomenology. Among these terms, the tensor one is known to impact nuclear structure properties, especially in neutron-rich nuclei. However, its effect has not been studied on the whole nuclear chart yet. 
\item[Purpose] The impact of terms corresponding to the tensor at the Hartree level, is studied for infinite nuclear matter as well as deformed nuclei, by developing new density-dependent functionals including these terms. In particular, we study in details the improvement such a term can bring to the description of specific nuclear observables.
\item[Methods] The framework of covariant energy density functional is used at the Hartree-Bogoliubov level. The free parameters of covariant functionals are optimized by combining Markov-Chain-Monte-Carlo and simplex algorithms. 
\item[Results] An improvement of the RMS binding energies, spin-orbit splittings and gaps is obtained over the nuclear chart, including axially deformed ones, when including tensors terms. Small modifications of the potential energy surface and densities are also found. 
In infinite matter, the Dirac mass is shifted to a larger value, in better agreement with experiments. 
\item[Conclusions] Taking into account additional terms corresponding to the tensor terms in the vector-isoscalar channel at the Hartree level, improves the description of nuclear properties, both in nuclei and in nuclear matter.
\end{description}
\end{abstract}

\maketitle

\section{Introduction}

The covariant Energy Density Functional (cEDF) approach achieved great success in describing finite nuclei and infinite nuclear matter properties \cite{vre05}.
The covariant formulation provides a natural mechanism for the appearance of central and spin-orbit (SO) parts of the interaction in terms of combinations of scalar and vector potentials. This allows to treat these terms on equal footing, in a more economical way.

The tensor force is of particular importance for the nucleon-nucleon interaction, first recognized to be responsible for the deuteron binding energy \cite{rar41} and non-zero electric quadrupole moment of the deuteron \cite{ger42}. Today, the impact of the tensor term has been studied in details for interactions, both covariant \cite{ruf88,lon07,typ20} or not \cite{sta77,dec80,les07}. 
It is expected that this term acts on the SO splitting between single-nucleon levels. Indeed, the latter mainly depends on the Dirac effective mass, which is linked to the scalar potential; introducing tensor terms increases the Dirac mass, while keeping reliable description of SO splittings.

In a covariant framework, the nucleon-nucleon interaction can be introduced by meson exchange and the tensor terms are defined as derivative terms in the vector-isoscalar ($\omega$) and vector-isovector ($\rho$) channels. 
Since derivative terms are the simplest terms to be added to a functional, tensor terms can also be considered as the next relevant contribution to an EDF based interaction.

Historically, the first appearance of explicit tensor couplings in RMF framework can be found in \cite{ruf88},  with non-linear coupling for the scalar-scalar degree of freedom in spherical nuclei. This study showed a negligible impact of the $\rho$ tensor coupling, while the $\omega$ one seemed to improve slightly the fit of the interaction, with an increased effective mass.
Many studies were then carried out to extend these calculations to the deformed case at the Hartree level \cite{jia91}.
Numbers of specific studies have been done to understand the effect of tensor terms on e.g. spin-orbit splittings \cite{ren95,fur98,jia15}, shell gaps \cite{wan13}, surface thickness \cite{jia05}, pseudo spin-orbit splitting \cite{lon07}, nuclear matter properties \cite{typ20}. 

The full treatment of the tensor term would require the inclusion of the Fock term. However, this precludes from making large scale calculations on the nuclear chart, due to the complexity of a tensor covariant Hartree-Fock approach. Indeed, a study of the interplay of the tensor terms together with pairing and deformation, in a covariant approach, is still lacking. This can be undertaken as the Hartree level, where the tensor terms rather acts as an extension of the functionnal than a full treatment of this term. Nevertheless, such a study can give hints of the behavior of the tensor effect over the nuclear chart. Moreover, a known problem with relativistic functionals is the low value of the effective Dirac mass $M^\star = M + S$, usually around $M^\star / M \approx 0.6$, instead of the empirically determined $M^\star /M \approx 0.75$. The inclusion of a tensor term allows to partially decouple the scalar and vector part of the interaction an should allow for a better description of the effective mass, by decreasing the value of the scalar potential.

In this work, new parametrizations of cEDF, with density dependent coupling constant, are introduced at the Hartree-Bogoliubov level.
The corresponding free parameters are optimized by means of least-square procedure with the use of both Markov-Chain-Monte-Carlo method \cite{nea93,for12} and simplex minimization \cite{pre86}. Constraints on binding energy, radii, spin-orbit splittings and level gaps, as well as several infinite matter constraints are considered.
The effect of tensor coupling at the Hartree level is studied in detail, in both spherical and axially deformed systems. 

This work is organized as follows : Sec. II introduces the theoretical framework, while Sec. III focuses on the fitting procedure. General results of the minimization process are given in Sec. IV, and proper applications to infinite matter, binding energies, radii, effect of deformation, SO splittings, gaps and densities are shown in Sec. V.

\section{Lagrangian and equations of motion}

We treat nuclei as collections of structure-less nucleons whose strong interactions are described in terms of (effective) mesons exchange, and with electromagnetic interactions mediated by photons. As spin-1/2 and isospin-1/2 fermions, nucleons are described by the isospin doublet 
\begin{equation}
    \Psi(x) = \begin{pmatrix}\psi_n(x)\\ \psi_p(x)\end{pmatrix}; \quad \overline\Psi(x) = \begin{pmatrix}\psi_n^\dagger(x)\gamma^0 & \psi_p^\dagger(x)\gamma^0\end{pmatrix}, 
\end{equation}
where $\psi_i(x)$ ($i=$neutron, proton) is a Dirac spinor field and $\gamma^{\mu=0}$ the corresponding Dirac matrix. The effective mesons and the photon are represented by a boson field $\Phi_b^{(J^\Pi,T)}(x)$ carrying the quantum numbers $(J^\Pi,T)$ (total angular momentum, parity and isospin). They dictate the behaviour of the bosons under Lorentz transformations and rotations in isospin space.

The equations of motion for the nucleonic and mesonic (and photonic) degrees of freedom are obtained from a covariant Lagrangian (density), which can be split into two sectors,
\begin{equation}
    \mathcal{L}=\mathcal{L}_{\text{NN}}+\mathcal{L}_{\text{bos}} .
\label{eq:Lag_tot}    
\end{equation}
The Lagrangian in the nucleon-nucleon (NN) sector, $\mathcal{L}_{\text{NN}}$, includes all the terms bilinear in the nucleon field $\Psi$, i.e.
\begin{equation}
    \mathcal{L}_{\text{NN}}=\overline{\Psi}\left(i \gamma^\mu\partial_\mu - M - g_b\Gamma \Phi_b^{(J^\Pi,T)} O_\tau \right)\Psi, 
\label{eq:Lag_N}    
\end{equation}
where M is the nucleon mass (we take $M=M_p=M_n$) and $g_b$ parametrizes the coupling between the boson (meson or photon) field $\Phi_b^{(J^\Pi,T)}(x)$ and the spinor bilinear $\left(\overline{\Psi}\Gamma O_\tau\Psi\right)(x)$. The coupling constants $g_b$, and in particular their density dependency, are discussed below in section \ref{sec:param}.
The matrix $\Gamma$ generically refers to one element of the set $\lbrace 1_4, \gamma^5,\gamma^\mu, \gamma^5\gamma^\mu,\sigma^{\mu\nu} \rbrace$, where $1_4$ is the four-by-four identity matrix, $\gamma^a$ ($a = 0,1,2,3,5$) are the Dirac matrices and $\sigma^{\mu\nu}\equiv-\frac{i}{4}\left[\gamma^{\mu},\gamma^{\nu}\right]$, forming a basis for the space of four-by-four complex matrices. The matrix $O_\tau\in\lbrace 1_2,\tau_i\rbrace$ ($i=1,2,3$) is either the two-by-two identity matrix or one of the three Pauli matrices $\tau_i$ in isospin space. All possible independent spinor bilinears can be formed with the $\Gamma$ and $O_\tau$ matrices, namely 
 \begin{itemize}
     \item[$\bullet$] five isoscalar bilinears :
     \begin{enumerate}
         \item[i)] $\overline{\Psi}1_4 1_2 \Psi$ (scalar),
         \item[ii)] $\overline{\Psi}\gamma^5 1_2 \Psi$ (pseudoscalar),
         \item[iii)] $\overline{\Psi}\gamma^\mu 1_2 \Psi$ (4-vector),
         \item[iv)] $\overline{\Psi}\gamma^5\gamma^\mu 1_2 \Psi$ (pseudo 4-vector),
         \item[v)] $\overline{\Psi}\sigma^{\mu\nu} 1_2 \Psi$ (rank-2 antisymmetric tensor),
     \end{enumerate}
          \item[$\bullet$] five isovector bilinears :
     \begin{enumerate}
         \item[i)] $\overline{\Psi}1_4 \vec{\tau} \Psi$ (scalar),
         \item[ii)] $\overline{\Psi}\gamma^5 \vec{\tau} \Psi$ (pseudoscalar),
         \item[iii)] $\overline{\Psi}\gamma^\mu \vec{\tau} \Psi$ (4-vector),
         \item[iv)] $\overline{\Psi}\gamma^5\gamma^\mu \vec{\tau} \Psi$ (pseudo 4-vector),
         \item[v)] $\overline{\Psi}\sigma^{\mu\nu} \vec{\tau} \Psi$ (rank-2 antisymmetric tensor),
     \end{enumerate}
 \end{itemize}
where we use arrows for isovectors. Hereafter, we omit the identity matrices $1_4$ and $1_2$. 

The spinor bilinears are coupled to bosonic fields $\Phi_b^{(J^\Pi,T)}(x)$ and derivatives thereof to eventually yield a scalar-isoscalar contribution to Lagrangian~\eqref{eq:Lag_N}. In addition to the electromagnetic coupling between protons
\begin{equation}
\mathcal{L}_{\text{NN}}^{\gamma}=\left[e\overline\Psi\gamma^{\mu}A_{\mu}\frac{1-\tau_3}{2}\Psi\right](x),    
\end{equation}
where $A_\mu$ is the electromagnetic 4-potential and $e$ the elementary proton charge, cEDFs only include the minimal set of nucleon(N)-meson couplings yielding a satisfactory description of nuclear bulk properties. A standard choice involves  (i) the $\sigma$, $\omega$ and $\rho$ (effective) mesons,  with quantum numbers respectively equal to $(0^+,0)$, $(1^-,0)$ and $(1^-,1)$, and therefore respectively represented by a scalar isoscalar field $\sigma(x)$, a 4-vector isoscalar field $\omega^\mu(x)$ and a 4-vector isovector field $\vec{\rho}^\mu(x)$ and (ii) the N-$\sigma$ scalar coupling as well as the N-$\omega$ and N-$\rho$ vector couplings, which respectively read
\begin{subequations}
\begin{align}
\mathcal{L}_{\text{NN}}^\sigma &=\left[g_\sigma \overline \Psi\sigma\Psi\right](x), \\
\mathcal{L}_{\text{NN}}^\omega &=\left[g_\omega \overline\Psi\gamma^\mu\omega_\mu\Psi\right](x), \\
\mathcal{L}_{\text{NN}}^\rho &=\left[g_\rho \overline\Psi\gamma^\mu\vec{\rho}_\mu\star\vec{\tau}\Psi\right](x),
\end{align}
\end{subequations}
where $\star$ refers to the scalar product in isospin space, while $g_i$ ($i=\sigma,\omega,\rho$) stands for the (density-dependent) N-$i$ coupling constant, to be adjusted. It should be noted that, in general, a pseudovector coupling between the nucleon and the pion is not considered when the cEDF is treated at the so-called relativistic mean field (RMF) level, where the exchange contributions are not computed explicitly. Indeed, if the reflection symmetry is preserved at the RMF level, the N-$\pi$ coupling yields a null direct contribution.

In this work, we enrich standard cEDFs by including the next simplest terms, i.e. the N-$\omega$ and N-$\rho$ (Lorentz) tensor couplings
\begin{equation}
    \mathcal{L}_{\text{NN}}^{\omega+\rho;T}=\left[\overline\Psi\sigma^{\mu\nu}\left(\frac{\Gamma_\omega^T}{2M}\Omega_{\mu\nu}+\frac{\Gamma_\rho^T}{2M} \vec{\mathcal{R}}_{\mu \nu} \star \vec{\tau}\right)\Psi\right](x),    
\end{equation}
where $\Gamma_i^T$ ($i=\omega,\rho$) stands for the (density-independent) N-$i$ tensor coupling constant, $\Omega_{\mu \nu}=\partial_\mu \omega_\nu - \partial_\nu \omega_\mu$ and $\vec{\mathcal R}_{\mu \nu}=\partial_\mu \vec\rho_\nu - \partial_\nu \vec\rho_\mu$ are the $\omega$ and $\rho$ field strength tensors, respectively.    

$\mathcal{L}_{\text{bos}}$ is the Lagrangian for the bosonic degrees of freedom, i.e. the meson fields and the electromagnetic 4-potential:
\begin{align}
    \mathcal{L_\text{bos}}  =& \frac{1}{2}\left(\partial_{\mu}\sigma\partial^{\mu}\sigma-m_{\sigma}^{2}\sigma^{2}\right)-\frac{1}{4}\left(\Omega_{\mu\nu}\Omega^{\mu\nu}-m_{\omega}^{2}\omega_{\mu}\omega^{\mu}\right) \nonumber\\
    &-\frac{1}{4}\left(\vec{\mathcal{R}}_{\mu\nu}\star\vec{\mathcal{R}}^{\mu\nu}-m_{\rho}^{2}\vec{\rho}_{\mu}\star\vec{\rho}^{\mu}\right)-\frac{1}{4}\left(F_{\mu\nu}F^{\mu\nu}\right),
    \label{eq:Lag_M}    
\end{align}
with $m_i$ ($i=\sigma,\omega,\rho$) the mass of the meson $i$ and $F_{\mu \nu}$ the electromagnetic field strength tensor.

Treating Lagrangian~\eqref{eq:Lag_tot} in the relativistic Hartree-Bogoliubov (RHB) approximation, eventually yields the equation of motion for a quasi-nucleon (in the Bogoliubov sense) in the quantum state $k$:
\begin{equation}
\begin{pmatrix}
    h_D(\boldsymbol{q})-\lambda & \Delta(\boldsymbol{q}) \\
    -\Delta^*(\boldsymbol{q}) & -h^*_D(\boldsymbol{q}) + \lambda
    \end{pmatrix}\begin{pmatrix}
    \mathcal{U}(\boldsymbol{q}) \\
    \mathcal{V}(\boldsymbol{q})
    \end{pmatrix}_k = E_k(\boldsymbol{q}) \begin{pmatrix}
    \mathcal{U}(\boldsymbol{q}) \\
    \mathcal{V}(\boldsymbol{q})
    \end{pmatrix}_k,  
\label{eq:eom}
\end{equation}
where $E_k$ stands for the eigenenergy of the quasiparticle (qp) $k$ while the Dirac spinors $\mathcal{U}_k$ and $\mathcal{V}_k$ contribute to the qp wavefunction. The chemical potential is called $\lambda$ and $\boldsymbol{q}$ collects a set of constrained collective coordinates (e.g. deformation parameters, pairing gap, etc.). 

It should be noted that other approximations are needed to obtain Eq.~\eqref{eq:eom}, among which the no-sea approximation where the Dirac sea of states with negative energies does not contribute to the densities and currents \cite{fur95}, or the omission of the time-dependence of the meson fields (it amounts to consider the exchange of mesons as an instantaneous process, which can be justified by their heavy mass compared to the typical relative momenta between the two interacting nucleons), or the preservation of time-reversal symmetry, with the consequence that only the time-like component of the 4-vector fields contribute (currents do not contribute). 

The fields $h_D$ and $\Delta$ are the RHB mean potential in the particle-hole and particle-particle channels respectively. More precisely, the single-nucleon Dirac Hamiltonian reads
\begin{align}
h_\text{D} =-i\boldsymbol{\alpha\cdot\nabla}+\beta \left[M^{\star}\left(\mathbf{r}\right) +\boldsymbol{\sigma ^0} \cdot \mathbf{T}(\mathbf{r})\right] +V\left(\mathbf{r}\right).
\label{eq:hd}    
\end{align}
In the last equation, bold symbols refer to 3-vector in real space and $\gras{\cdot}$ indicates the corresponding scalar product, $\gras{\alpha}=\gamma^0\gras{\gamma}$, $\gras{\gamma}$ is a 3-vector with components $\gamma^i$ ($i=1,2,3$), $\beta=\gamma^0$ and $\gras{\sigma}^0$ is a 3-vector with components $\sigma^{0i}$ ($i=1,2,3$). $M^*$ stands for the Dirac effective mass
\begin{equation}
    M^\star(\gras{r}) = M + S(\gras{r}),
\end{equation}
involving the nucleon scalar self-energy
\begin{equation}
S =  g_\sigma \sigma.
\end{equation}
$V(\gras{r})$ is the nucleon vector (time-like component) self-energy
\begin{equation}
    V= g_\omega \omega + \tau_3 g_\rho \rho + e\frac{1+\tau_{3}}{2} A + V^\text{(R)},
\end{equation}
where we have set $\omega(\gras{r})\equiv\omega^0(\gras{r})$ and $\rho(\gras{r})\equiv \rho^0_3(\gras{r})$ (only the time-like component of 4-vectors and the third component of isovectors contribute due to the set of approximations made, as discussed above) and where $V^{(R)}$ is the so-called rearrangement term
\begin{equation}
    V^\text{(R)} = \frac{dg_\sigma}{d\rho_V} \rho_S\sigma +\frac{dg_\omega}{d\rho_V} \rho_V\omega + \tau_3 \frac{dg_\rho}{d\rho_V} \rho_{TV}\rho. 
\end{equation}
This equation involves the scalar, vector and isovector densities 
\begin{subequations}
\begin{align}
    \rho_S &= \mathcal{V}^*\beta \mathcal{V}^T, \\
    \rho_V &= \mathcal{V}^*\mathcal{V}^T, \\
    \rho_{TV} &= \mathcal{V}^*\tau_3\mathcal{V}^T.
\end{align}
\end{subequations}
The new contribution $\gras{\sigma^0\cdot T}(\gras{r})$ coming from the tensor couplings modifies the Dirac effective mass. It reads
\begin{equation}
    \mathbf{T} = - \frac{\Gamma_\omega^T}{M} \boldsymbol{\nabla} \omega  - \tau _3 \frac{\Gamma_\rho^T}{M} \boldsymbol{\nabla} \rho.
\end{equation}

At the RHB level, the bosonic degrees of freedom's equations of motion reduce to Klein-Gordon equations:
\begin{subequations}
\begin{align}
    \left(-\gras{\nabla}^2+ m_\sigma ^2\right) \sigma &= -g_\sigma \rho_S, \\
    \left(-\gras{\nabla}^2+ m_\omega ^2\right)\omega &= g_\omega \rho_V + \frac{\Gamma_\omega^T}{M} \boldsymbol{\nabla} \cdot \gras{j}_{T\omega},\\
    \left(-\gras{\nabla}^2+ m_\rho ^2\right)\rho &= g_\rho \rho_{TV} + \frac{\Gamma_\rho^T}{M} \boldsymbol{\nabla} \cdot \gras{j}_{T\rho},\\
    -\gras{\nabla}^2 A &= e\rho_p,
\end{align}
\end{subequations}
with the isoscalar and isovector tensor currents
\begin{subequations}
\begin{align}
\gras{j}_{T\omega} &= \mathcal{V}^* \beta\gras{\sigma}^0 \mathcal{V}^T, \\    
\gras{j}_{T\rho} &= \mathcal{V}^* \beta\gras{\sigma}^0\tau_3 \mathcal{V}^T,    
\end{align}
\end{subequations}
and the proton density 
\begin{eqnarray}
\rho_p\equiv \mathcal{V}^*\frac{1-\tau_3}{2}\mathcal{V}^T.    
\end{eqnarray}

Lagrangian~\eqref{eq:Lag_tot} only contributes to the particle-hole channel. It is complemented
by a separable pairing force in momentum space~\cite{dug04,tia09} 
$\displaystyle \langle k | V^{^1S_0}|k^\prime\rangle = -G p(k)p(k^\prime)$ in the particle-particle channel. By assuming a simple Gaussian ansatz
$p(k) = e^{-a^2k^2}$, the two parameters $G$ and $a$
are typically adjusted to reproduce the density dependence of the pairing gap
at the Fermi surface, obtained in nuclear matter with the Gogny D1S parametrization~\cite{ber91}.

The coupled nucleonic and bosonic equations of motion are expanded in a harmonic oscillator basis, for which details can be found in App.\ref{app:sph} and App.\ref{app:ax}. Solving them, yields the qp energies and wavefunctions, from which one can compute nuclear observables (e.g. binding energies and radii) as well as the canonical single-particle spectrum.

The total energy is computed as
\begin{equation}
    E = E_\text{part} - E_\sigma - E_\omega - E_\rho - E_C - E_R + E_\text{pair} + E_\text{CM}
\end{equation}
where $E_\text{part}$ represents the sum over the particles energies obtained by diagonalizing the Hartree Hamiltonian. The $E_\phi$ ($\phi = \sigma, \omega, \rho$) represent the mesonic field energies, $E_C$ the energy of the Coulomb field, $E_R$ the rearrangement energy, $E_\text{pair}$ the pairing energy and $E_\text{CM}$ the centre of mass correction. The latter one is computed as $E_\text{CM} = \left< P ^2\right>$/2M. The tensor contributions are directly taken into account in the mesonic field energies $E_\omega$ and $E\rho$.
The RMS radius $r_\text{RMS}$ and charge radius $r_C$ are defined as 
\begin{align}
    r_{RMS} &= \int d^3 r \rho (r) r^2 \\
    r_C     &= \sqrt{r_p ^2 + 0.64}
\end{align}
where the factor $0.64$ accounts for the finite size of the proton.

Finally, it has been checked that all the results are stable with $N=18$ harmonic oscillator shells, which are then considered for all the calculations.

\section{Optimization of the energy density functionals}

\subsection{Free parameters\label{sec:param}}

The cEDF involves the N-meson coupling constants as well as the nucleon and mesons masses as free parameters. The scalar and vector coupling constants $g_i = g_i(\rho_V(\gras{r}))$ ($i=\sigma, \omega, \rho$) are taken as explicit functions of the vector density \cite{typ99} :
\begin{equation}
    g_i(\rho_V(\gras{r})) = \Gamma_i h_i(\xi),\qquad i=\sigma, \omega, \rho,    
\end{equation}
where 
\begin{subequations}
\begin{align}
    \Gamma_i &\equiv g_i(\rho_\text{sat}), \\
    \xi &\equiv \frac{\rho_V(\gras{r})}{\rho_\text{sat}}.
\end{align}
\end{subequations}
The $h$ functions read
\begin{equation}
    h_i(\xi) \equiv a_{i}\frac{1+b_{i}\left(\xi+d_{i}\right)^{2}}{1+c_{i}\left(\xi+d_{i}\right)^{2}}, \qquad i=\sigma, \omega,
\end{equation}
in the isoscalar channel ($i=\sigma, \omega$) and  
\begin{equation}
    h_\rho(\xi) \equiv \text{exp}\left[-a_\rho\left(x-1\right)\right],
\end{equation}
in the isovector channel. 

The parameters entering the definition of the density-dependent coupling constants are not independent variables. A first obvious constraint requires that $h_i(1)=1$, yielding the relation
\begin{equation}
a_{i}=\frac{1+c_{i}\left(1+d_{i}\right)^{2}}{1+b_{i}\left(1+d_{i}\right)^{2}},\qquad i=\sigma, \omega.    
\end{equation}
An additional constraint can be imposed on the second derivative of the $h$ functions in order to ensure that the rearrangement contributions do not diverge at zero density: $h_i^{\prime\prime}(0) = 0$ ($i=\sigma,\omega$), leading to  
\begin{equation}
    c_{i}=\frac{1}{3d_{i}^{2}},\qquad i=\sigma, \omega.    
\end{equation}
In summary, the coupling constants use 11 free parameters, i.e. the 6 parameters $\Gamma_i$, $b_i$ and $d_i$ ($i=\sigma,\omega$) in the isoscalar channel, the 2 parameters $\Gamma_\rho$ and $a_\rho$ in the isovector channel, the saturation density $\rho_\text{sat}$ as well as the 2 density-independent tensor couplings $\Gamma_i^T$ ($i=\omega,\rho$).

The nucleon mass is set to $M=M_n=M_p=939$ MeV, the $\omega$ and $\rho$ meson masses are fixed to their observed value in free space, i.e. $m_\omega=783$ MeV and $m_\rho=763$ MeV, while the $\sigma$ meson mass remains a free parameter to be adjusted. 

In order to reliably assess the impact of the tensor couplings on nuclear properties at the RHB level, two new parametrizations will be derived, using the same fitting protocol: one for a cEDF without the tensor couplings (10 free parameters) and one for a cEDF with the $N-\omega$ and $N-\rho$ tensor couplings (12 free parameters).

\subsection{Experimental dataset}

Before discussing the fitting protocol, we first present the pool of empirical data selected to calibrate the cEDFs. The database contains both infinite homogeneous nuclear matter and finite nuclei properties $\mathcal{O}_i$, each with an uncertainty $\Delta\mathcal{O}_i$, enabling a measure of the quality of the optimal parametrization via a sensitivity analysis. 

A first set of constraints comes from the properties of infinite nuclear matter, namely the isoscalar ones: energy per particle of symmetric nuclear matter at equilibrium $E_0$, incompressibility $K_0$, and the isovector ones: symmetry energy coefficient at saturation density $J$, and $K_\text{sym}$. These parameters characterize the nuclear equation of state around saturation density $\rho_\text{sat}$ and isospin asymmetry $\delta\equiv(\rho_n-\rho_p)/\rho=0$ (see App.\ref{app:EoS}). 
The selected parameters are reported in Table~\ref{tab:constraints_infinite}, with their considered values -stemming from (model-dependent) extrapolations of measured properties in finite nuclei- and associated accuracy to which these values should be reproduced in the fit. The targets values are chosen from empirical results \cite{mar18} and previous covariant parametrizations.

\begin{table}
    \centering
    \begin{tabular}{ P{15mm} P{15mm} P{15mm} P{15mm} P{15mm} }
        \hline
        \hline
        & $\rho _{sat}$ (fm$^{-3}$) & $E_0$ (MeV) & $K_0$ (MeV) & $J$ (MeV) \\
        \hline
        $\mathcal O$ & 0.152 & -16.15 & 235 & 31.5 \\
        $\Delta \mathcal O$ & 0.03 & 0.15 & 30 & 3 \\
        \hline
        \hline
    \end{tabular}
    \caption{Infinite nuclear matter pseudo-data considered in the fit, with their allowed uncertainties.}
    \label{tab:constraints_infinite}
\end{table}

In addition to the infinite nuclear matter pseudo-data, the finite nuclei properties embrace data for 12 spherical nuclei, presented in Table~\ref{tab:constraints}. It should be noted that the numerical cost associated to the optimization algorithm makes it challenging to include deformed nuclei in the database, generally increasing the computation time by one or two order of magnitude. 
The selected experimental dataset involves binding energies (BE), charge radii, differences between neutron and proton RMS. radii and single-particle level splittings. 

Other constraints were tested but did not impact the interaction.
Among the observables tested were, for example
\begin{itemize}
    \item a constraint on neutron matter, as indicated in the fit procedure of Gogny D1N interaction, which helps to reduce the effects of isotopic drifts \cite{cha08}
    \item a constraint on the effective vector mass $M_V = M - V$ that would allow a better reproduction of the experimental results \cite{nik03}
\end{itemize}

\begin{table*}
    \centering
    \begin{tabular}{ P{15mm} P{19mm} P{18mm} P{18mm}  P{20mm} P{50mm} P{20mm} }
    \hline \hline
    Nucleus & BE/A (MeV) & $r_\text{charge}$ (fm) & $r_n - r_p$ (fm) & $\Delta _\text{so}$ (MeV) & Gaps (MeV) & Comments\\
    
    \hline
 
    $^{16}$O & 7.976 & 2.730 & & & 11.61 (p 1d5/2 - 1p1/2) & Z=8 gaps \\ & & & & & 11.55 (n 1d5/2 - 1p1/2) & N=8 gap\\

    $^{22}$O & 7.364 &  &  & & 10.25 (p 1d5/2 - 1p1/2) & Z=8 gap\\
    
    $^{40}$Ca & 8.551 & 3.485 & & & 07.25 (p 1f7/2 - 1d3/2 (2s1/2)) & Z=20 gap\\ & & & & & 07.30 (p 1f7/2 - 1d3/2) & N=20 gap\\
    
    $^{48}$Ca & 8.667 & 3.484 & & & 06.55 (p 1f7/2 - 1d3/2 (2s1/2)) & Z=20 gap\\

    $^{54}$Ca & 8.247 & & & & 06.24 (p 1f7/2 - 1d3/2 (2s1/2)) & Z=20 gap\\
    
    $^{56}$Ni & 8.642 & & & &  06.48 (p 2p3/2 - 1f7/2) & Z=28 gap\\ & & & & & 06.36 (n 2p3/2 - 1f7/2) & N=28 gap\\
    
    $^{60}$Ni & 8.780 & & & &  04.74 (p 2p3/2 - 1f7/2) & Z=28 gap\\
    
    $^{66}$Ni & 8.739 & & & &  05.53 (p 2p3/2 - 1f7/2) & Z=28 gap\\ 
    
    $^{90}$Zr & 8.710 & 4.272 & & & &\\
    
    $^{116}$Sn & 8.523 & 4.626 & 0.120  & & &\\
    
    $^{124}$Sn & 8.467 & 4.674 & 0.190 & & &\\
    
    $^{132}$Sn & 8.354 & 4.709 & & 01.48 (2d) & &\\ & & & & 06.14 (1g) & & \\
    
    \hline 
    $\Delta \mathcal O _i$ & 0.03 & 0.01 & 0.01 & 0.2 & 0.2 &\\
    
    \hline \hline
    
    \end{tabular}
    \caption{Constraints set on finite nuclei.}
    \label{tab:constraints}
\end{table*}

\subsection{Optimization protocol}

The vector of free parameters to be fitted, characterizing cEDFs without and with tensor couplings, are called 
\begin{subequations}
\begin{align}
    \mathfrak{p}_{\cancel{T}} &= \left(m_\sigma,\Gamma_\sigma,b_\sigma,d_\sigma,\Gamma_\omega,b_\omega,d_\omega,\Gamma_\rho,a_\rho,\rho_\text{sat}\right), \\
    \mathfrak{p}_T &= \left(\mathfrak{p}_{\cancel{T}},\Gamma_\omega^T,\Gamma_\rho^T\right),
\end{align}
\end{subequations}
respectively, and are generically referred as $\mathfrak{p}$ when no distinction between them is necessary. The objective function, in the optimization protocol, is chosen under the form of a weighted sum of squared differences
\begin{equation}
    \chi^{2}(\mathfrak{p})=\sum_{i}\left(\frac{\mathcal{O}_{i}^{th}(\mathfrak{p})-\mathcal{O}_{i}^{exp}}{\Delta \mathcal{O}_{i}}\right)^{2},
    \label{eq:chi2}
\end{equation}
with $\mathcal{O}_i^{th}$ (resp. $\mathcal{O}_i^{exp}$) the simulated (resp. experimental) value for one element $\mathcal{O}_i$ of the selected dataset, and $\Delta\mathcal{O}_i$ the corresponding estimated error through which one is able to balance the contributions from the different categories of data being simultaneously fitted. 

The search of a minimum for the objective function~\eqref{eq:chi2} over the domain of the parameter set is performed through two derivative-free optimizers, (i) the Nelder-Mead (NM) pattern search algorithm~\cite{pre86}, sometimes coined simplex method, which offers a great compromise between simplicity and efficiently, but from which a detailed analysis of the correlations in the parameter set is not accessible, and (ii) the Markov Chain Monte Carlo (MCMC) algorithm based on Metropolis-Hastings method \cite{for12} which provides a better exploration of the parameter space and gives access to the correlations among them, however for a larger computational cost. 

\begin{figure}
    \centering
    \includegraphics[width=1\linewidth]{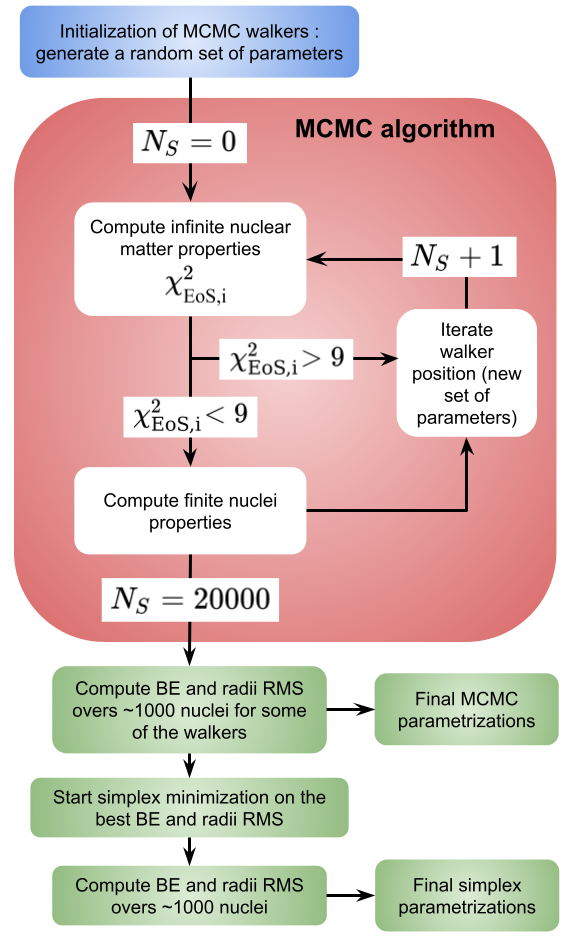}
    \caption{Schematic view of the different steps during the minimization procedure. 
    }
    \label{fig:flow_minim}
\end{figure}

The optimization algorithm proceeds as follows (see Fig.~\ref{fig:flow_minim}):
\begin{enumerate}
    \item A parameter set $\mathfrak{p}^0$ is randomly chosen, corresponding to an initial position of the MCMC walkers.
    \item The properties of infinite nuclear matter reported in  Table~\ref{tab:constraints_infinite} are computed for the set $\mathfrak{p}^0$ and used to construct the corresponding contributions to the objective function~\eqref{eq:chi2} $\chi^2_{\text{EoS},i}\left(\mathfrak{p}^0\right)$ ($i=\rho_\text{sat},E_0,K,J$).  
    \item If $\chi^2_{\text{EoS},i}\left(\mathfrak{p}^0\right)>9$, the set $\mathfrak{p}^0$ is rejected and the walkers position is iterated, yielding a new set $\mathfrak{p}^1$. Steps 2 and 3 are repeated until $\chi^2_{\text{EoS},i}\left(\mathfrak{p}^{\text{EoS ok};1}\right)<9$ for a parameter set $\mathfrak{p}^{\text{EoS ok};1}$.
    \item The properties of finite nuclei reported in Table~\ref{tab:constraints} are computed for the parameter set $\mathfrak{p}^{\text{EoS ok};1}$, as well as the corresponding total objective function~\eqref{eq:chi2}.  
    \item Steps 2 to 4 are repeated 20000 times after iterating on the walkers position, yielding 20000 parameter sets $\mathfrak{p}^{\text{EoS ok};j}$ with their corresponding $\chi_j^2$.
    \item The binding energy and radius of 828 deformed nuclei are determined for $N_\text{NM}$ of the 20000 parameter sets $\mathfrak{p}^{\text{EoS ok}; j}$. The corresponding RMS deviations $\delta_B^2(\mathfrak{p}^{\text{EoS ok};j}) = \sum_{k=1}^{828}\left(B_k^\text{th}(\mathfrak{p}^{\text{EoS ok};j})-B_k^\text{exp}\right)^{2}$ and $\delta_r^2(\mathfrak{p}^{\text{EoS ok};j}) = \sum_{k=1}^{828}\left(r_k^\text{th}(\mathfrak{p}^{\text{EoS ok};j})-r_k^\text{exp}\right)^{2}$ are then computed. 
    \item The $N_\text{NM}$ parameter sets $\mathfrak{p}^{\text{MCMC} \delta_\text{min};n}$ among the 20000 $\mathfrak{p}^{\text{EoS ok}; j}$ with the smallest r.m.s deviations are kept. A NM optimization is performed on each of these $N_\text{NM}$ parameter sets with respect to the objective function~\eqref{eq:chi2}, yielding the converged parameter sets $\mathfrak{p}^{\text{MCMC-NM};n}$, with their corresponding $\chi_n^2$.     
    \item Step 6 is repeated for each of the parameter sets $\mathfrak{p}^{\text{MCMC-NM};n}$. Finally, the parameter set $\mathfrak{p}^{\text{MCMC-NM};\delta_\text{min}}$ yielding the smallest RMS deviations is selected.
\end{enumerate}
$N_\text{NM}$ represent the number of sets tested with a NM algorithm among the most converged MCMC walkers. In this study, $N_\text{NM} = 20$.

\subsection{Results}

The MCMC simulations have been performed on the full parameter space, in order to monitor the different correlations which link the parameters of the functional. Each MCMC simulations ran with $N_w = 200$ walkers for $N_s = 2 \times 10^4$ steps, each leading to a total of $\sim 5.10^7$ converged spherical RHB calculations.

\begin{figure}
    \centering
    \includegraphics[width=1\linewidth]{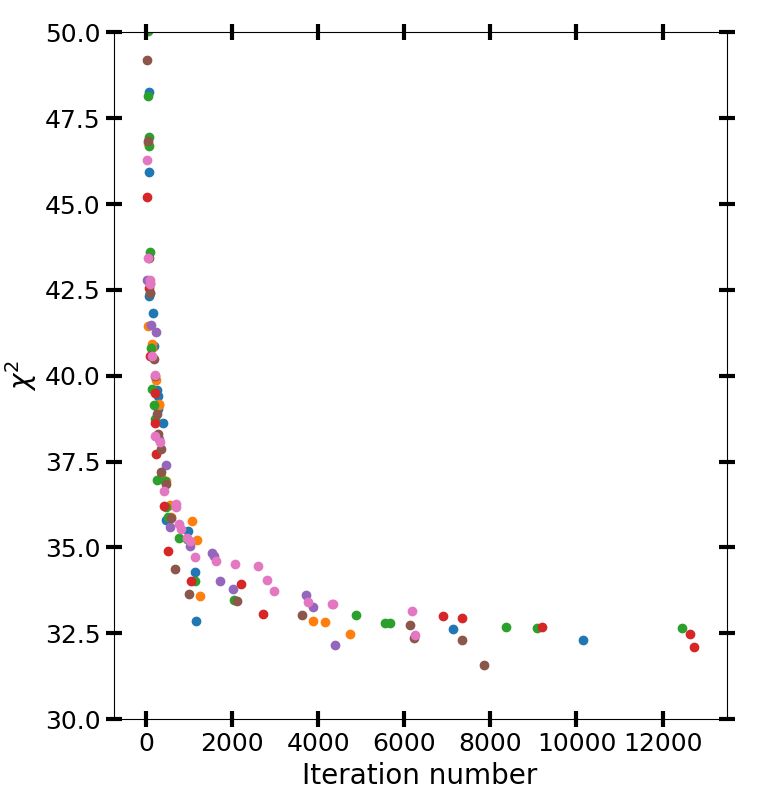}
    \caption{Value of $\chi ^2$ as a function of the iteration number for 7 of the 200 walkers represented by the different colors.}
    \label{fig:MCMC_convergence}
\end{figure}

The MCMC calculations converged to $\chi ^2 \sim 32$ as shown in Fig.~\ref{fig:MCMC_convergence}. Few lower minima can be found later on in the process, but as explained below, a lower $\chi ^2$ value is not systematically correlated to a lower binding energy (or radius) RMS over the nuclear chart.

It has been checked that results are not walkers dependent, in the sense that all the most converged calculations of a walker are similar to the ones of another walker. This check is important to ensure that the ending point of each walker is independent of its initial position (which is randomly chosen for the $N_w$ walkers).

The final choice of the functional remains a very subjective task, since many parameter sets reach identical values of $\chi ^2$ after thousands of steps. This is due to the fact that the parameters are strongly correlated between them, and a small modification of a single parameter can be balanced in many different ways. 

Another issue previously discussed also lies in the fact that a given parametrization, that has a better $\chi ^2$ than another one, may not give better results when applied to the entire nuclear chart. This problem mostly originates from the low number of nuclei included in the fitting procedure.
In order to choose the best parametrization, a subspace of the most converged (lowest $\chi ^2$) is picked. For each of these sets, a complete calculation of 828 nuclei is performed with an axial code. The binding energies and the charge radii are then compared to experimental values, leading to different RMS energies and radii for all the sets.

Another approach has also been tested where the most converged MCMC calculations were given as a starting point to a simplex minimization code. As expected, the simplex code allowed to find configurations that lead to smaller values of $\chi ^2$. These new "best fitted" parametrizations have also been tested over the nuclear chart and compared to the other ones.

\begin{figure}
    \centering
    \includegraphics[width=1\linewidth]{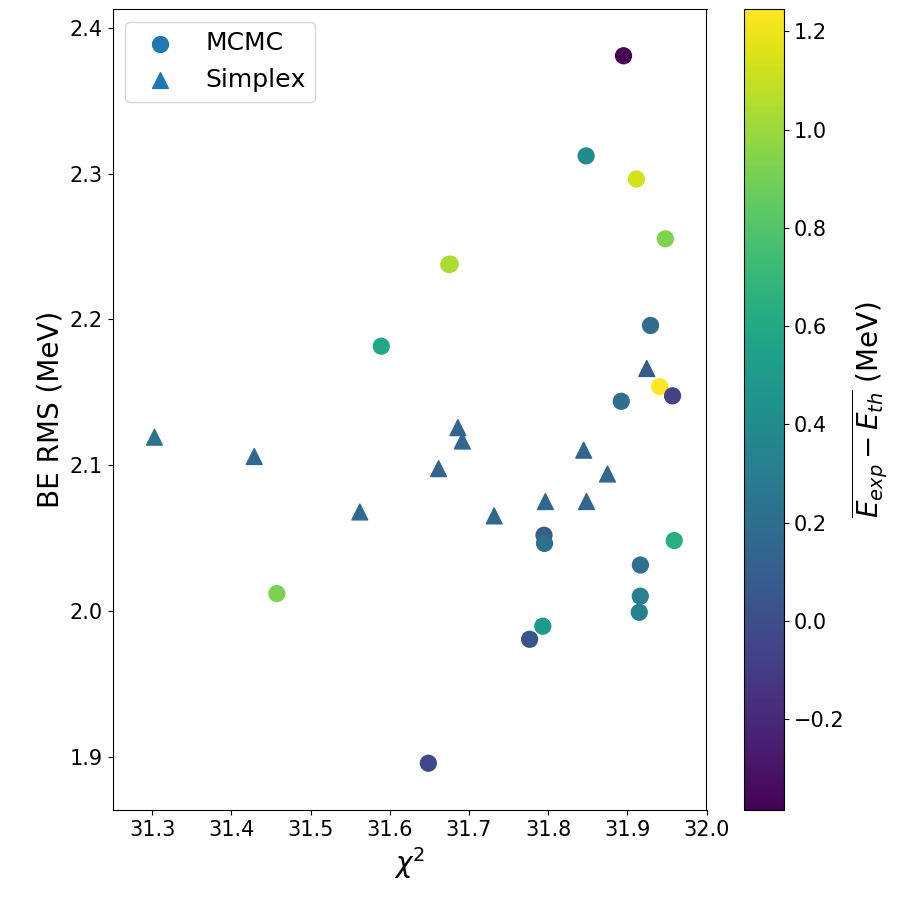}
    \caption{Binding energies RMS values on 828 nuclei as a function of $\chi ^2$. Dots represent the result of MCMC sampling while triangles represent the results of simplex minimization. The colorbar shows the mean value of $E_{exp} - E_{th}$. }
    \label{fig:chi2_RMS}
\end{figure}

In order to show that a lower $\chi ^2$ value does not necessarily correspond to lower RMS values, Fig.~\ref{fig:chi2_RMS} shows the binding energy RMS as a function of $\chi ^2$, for the different parametrizations. 
It should be noted that, sometimes, large mean value of $E_\text{exp}-E_\text{th}$ are obtained, leading to poor values of BE RMS (seen in the colorscale of Fig.~\ref{fig:chi2_RMS}). 
It should be noted that the parametrizations obtained with simplex minimization lead to quasi-constant BE RMS ($\sim 2.1$ MeV). This might show that all the simplex minimization ended up in equivalent local minima (due to correlation between parameters).
This is not the case for MCMC sampling, which explores much more of the parameter space and end up in non-equivalent local minima.

It is interesting to notice a weak correlation between the BE RMS and the $\chi ^2$ values for the MCMC sampling : a low value of $\chi ^2$ seems to be related with a low value of BE RMS. However, the statistical dispersion around this trend remains important and this criterion can not be used to determine the best fitted interaction.

Finally, the set with the lowest BE RMS (1.89 MeV) is selected. The values of these parameters set are given in Tab.\ref{tab:parameters}. 
However, it is important to notice that this choice remains subjective.

\begin{table}
    \centering
    \begin{tabular}{  P{13mm} P{15mm} P{26mm} P{26mm} }
    \hline \hline
    Param. & DD-ME2 & DD-MEV & DD-MEVT \\
    
    \hline
    $m_\sigma$ & 550.1238 & \bf{544.8503} & \bf{530.3956} \\
    (MeV) & & 543.8758 (1.0387) & 528.483 (2.4997) \\
    $\Gamma _\sigma$ & 10.5396 & \bf{10.2600} & \bf{9.2652} \\
     & & 10.168 (0.0686) & 9.141 (0.1428) \\
    $\Gamma _\omega$  & 13.0189 & \bf{12.7563} & \bf{11.6465} \\
     & & 12.6512 (0.102) & 11.5027 (0.1661)\\
    $\Gamma _\rho$ & 3.6836 & \bf{3.6709} & \bf{3.7971} \\
     & & 3.5827 (0.0648) & 3.7957 (0.0583)\\
    $b_\sigma$ & 1.0943 & \bf{1.4449} & \bf{2.0162} \\
     & & 1.4982 (0.1208) & 2.1246 (0.1487)\\
    $c_\sigma$ & 1.7057 & \bf{2.0619} & \bf{2.4924} \\
     & & 2.0708 (0.1296) & 2.5474 (0.1487)\\
    $b_\omega$ & 0.9240 & \bf{ 0.9828} & \bf{1.4375} \\
     & & 0.8793 (0.0735) & 1.5615 (0.1952)\\
    $c_\omega$ & 1.4620 & \bf{1.3672} & \bf{1.5696} \\
     & & 1.1762 (0.1175) & 1.6279 (0.2238)\\
    $a_{tv}$ & 0.5647 & \bf{0.5856} & \bf{0.5777} \\
     & &  0.6708 (0.0424) & 0.5446 (0.0387) \\
    $\rho _{sat}$ & 0.1520 & \bf{0.1505} & \bf{0.1516} \\
    (fm$^{-3}$) & & 0.1525 (0.0020) & 0.1517 (0.0023)\\
    $\Gamma ^T _\omega$ & 0.0 & \bf{0.0} & \bf{1.5496}\\
     & & & 1.7564 (0.2341)\\
    $\Gamma ^T _\rho$ & 0.0 & \bf{0.0} & \bf{-2.3243} \\
     & & & 0.2127 (2.0169)\\
    \hline \hline
    
    \end{tabular}
    \caption{Parameters values, shown on the first line of each cell (in bold). Mean values of MCMC sampling and associated uncertainties are shown on the second line of each cell. They are calculated using the 1000 most converged MCMC results.}
    \label{tab:parameters}
\end{table}

\section{Sensitivity analysis}
 
\subsection{Parameters sampling}
\begin{figure*}
    \centering
    \includegraphics[width=1\linewidth]{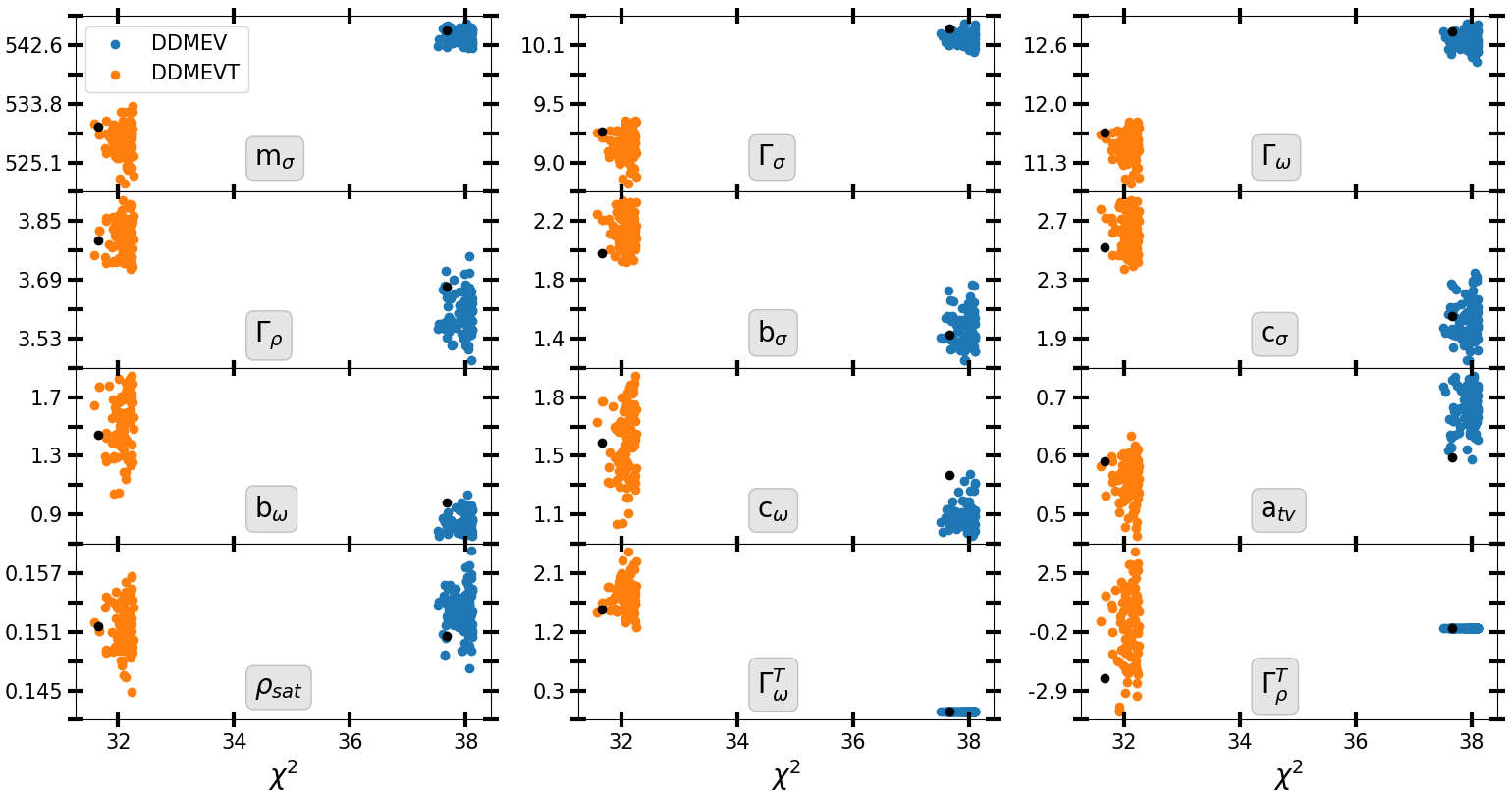}
    \caption{Parameter values of some of the most converged MCMC calculations for DD-MEV (blue) and DD-MEVT (orange). The black dots represent the parametrization that is used in the following sections for applications (see text for more details). The last two plots show the tensor coupling constants, explaining the constant 0 values of the DD-MEV parametrization for these parameters.}
    \label{fig:Para_diff_int}
\end{figure*}

It is relevant to first plot the values of some of the most converged calculations, for both interactions: DD-MEV (without the tensor term) and DD-MEVT (with the tensor term). Fig.~\ref{fig:Para_diff_int} displays the corresponding parameters values.

It is first interesting to notice that all the parameters have been shifted (except for the saturation density) between DD-MEV and DD-MEVT interactions, meaning that a fitting procedure, focusing only on the tensor d.o.f. on top of an older fitted interaction, would certainly not lead to a satisfactory interaction.

Another general behavior holds in the dispersion of the distribution. The interactions involving the new tensor terms exhibit larger dispersion for almost all the parameters. This might be linked to the addition of free parameters, and hence to some freedom in the fitting procedure.

An important result of Fig.~\ref{fig:Para_diff_int} is that the $\rho$ tensor part of the interaction is not constrained by our set (Tab.\ref{tab:constraints}). 
Indeed, this parameter takes value between -2 and 2 and does not seem to be correlated to any of the other parameters. 
Hence, the vector isovector tensor part of the interaction does not bring any physical content at the RHB level, or at least not on the observables that have been used here.

\subsection{Correlation and density dependence}
\begin{figure}
    \centering
    \includegraphics[width=1\linewidth]{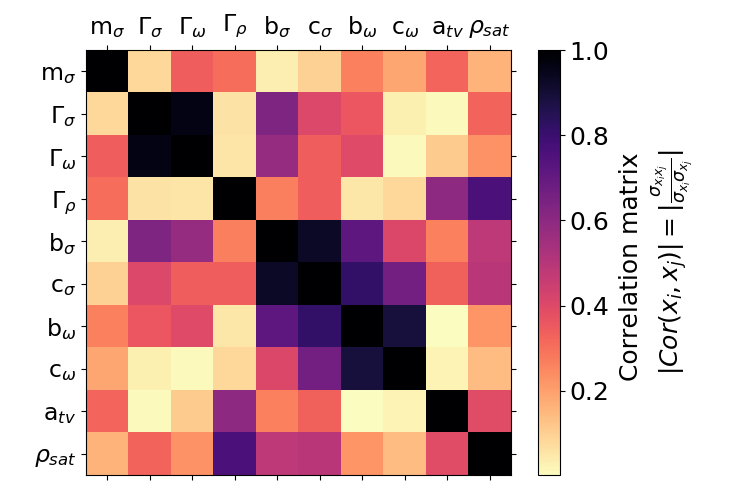}
    \caption{Correlation matrix obtained with the MCMC calculation for DD-MEV interaction.}
    \label{fig:Correlation_DDV}
\end{figure}

\begin{figure}
    \centering
    \includegraphics[width=1\linewidth]{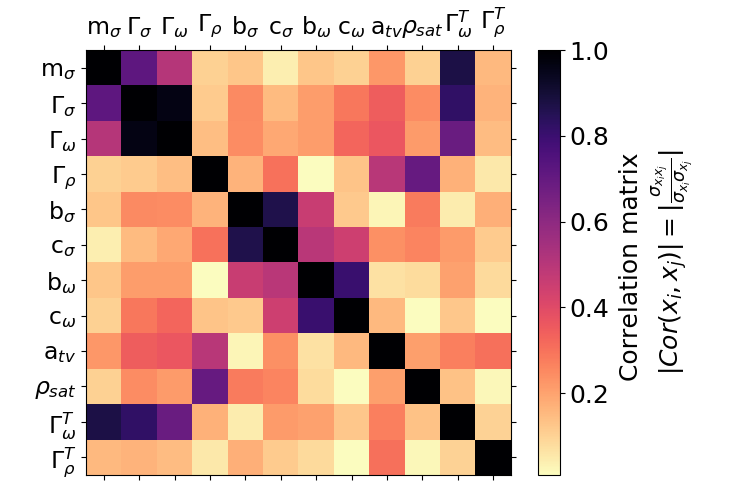}
    \caption{Correlation matrix obtained with the MCMC calculation for DD-MEVT interaction.}
    \label{fig:Correlation_DDVT}
\end{figure}

In order to understand the differences between the DD-MEV and DD-MEVT interactions, it is relevant to plot the correlation matrices. This is done in Fig.~\ref{fig:Correlation_DDV} and Fig.~\ref{fig:Correlation_DDVT}, where these matrices are defined as
\begin{equation}
    M_\text{corr} \left(x_{i},x_{j}\right) = \frac{\sigma _{x_i x_j}}{\sigma _{x_i} \sigma _{x_j}}
\end{equation}
where $\sigma _{x_i x_j}$ holds for the covariance matrix between the variable $x_i$ and $x_j$ and $\sigma _{x_i}$ is a shortcut for $\sigma _{x_i x_i}$. Here, we only focus on the absolute value of these matrices.

The first result to be noticed is the very strong correlation of the $m_\sigma$, $\Gamma _\sigma$ and $\Gamma _\omega$ variables for the DD-MEVT interaction. These three variables mainly encode the binding energy of the system and hence, a small variation in one of these variable can be compensated by the modification of the two others. 
It should be noted that $\Gamma ^T _\omega$ is also strongly correlated to these parameters as well, showing the importance of the vector isoscalar d.o.f. in encoding binding energy properties.

It should also be noted that in the case of DD-MEV interaction, the mass $m_\sigma$ is much less correlated to the other parameters. This shows that the presence of the $\omega$ tensor terms allows to introduce much more correlations in the isoscalar part of the interaction.

The second set of strongly correlated parameters are the ones encoding the density dependence of the isoscalar part of the interaction, namely $b_{\sigma,\omega}$ and $c_{\sigma,\omega}$. For these parameters, this trend is followed by both DD-MEV and DD-MEVT interactions.

The parameters encoding the isovector parts $\Gamma _\rho$ and $a_{tv}$ are also very correlated. However, a strong correlation between the saturation density and $\Gamma _\rho$ is also visible. This might be explained by the exponential form of the density dependence of the isovector sector, which enforce the importance of this parameter.

In the case of $\rho$ tensor term, it does not seem to be correlated to any variable. This is coherent with the previous result (Fig.~\ref{fig:Para_diff_int}) concerning its evolution with $\chi ^2$ : it is not constrained by our set or almost don't contribute.

As expected, the different shifts shown in Fig.~\ref{fig:Para_diff_int} are related to the correlation matrix previously introduced. In particular, the behavior of the IS channel shows decreasing values of the  $m_\sigma$, $\Gamma _\sigma$ and $\Gamma _\omega$ parameters when the tensor effect is considered w.r.t. to DD-ME2-like interaction, while $\Gamma ^T _\omega$ takes positive value.

\begin{figure}
    \centering
    \includegraphics[width=1\linewidth]{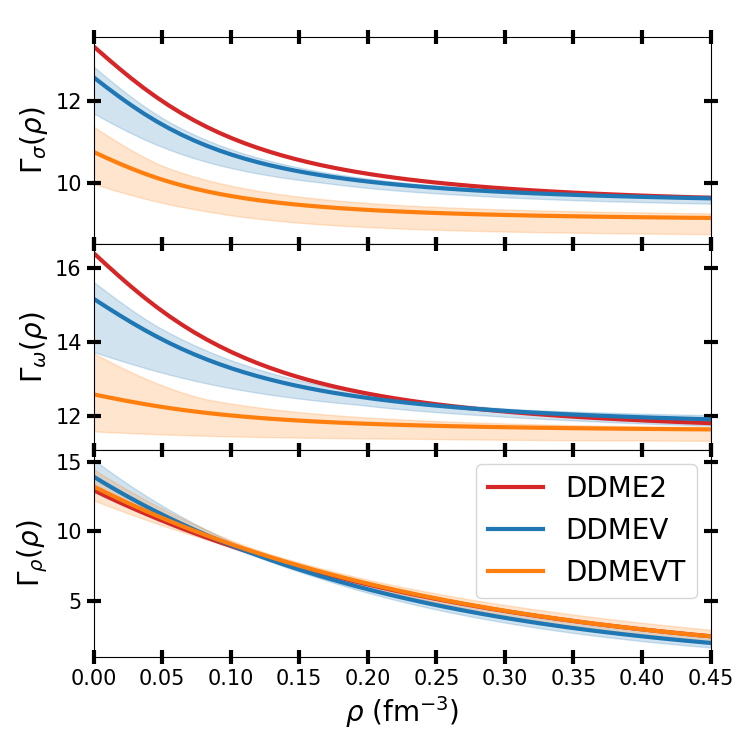}
    \caption{Evolution of the density dependent coupling constants $\Gamma _\sigma$, $\Gamma _\omega$ and $\Gamma _\rho$ as a function of the density $\rho$. Three parametrizations are shown : DD-ME2, DD-MEV and DD-MEVT. The shaded areas represent the standard deviations (see the text for more details).}
    \label{fig:Compare_DD}
\end{figure}

A less trivial shift occurs for the $b_i$ and $c_i$ parameters, which are linked to the parametrization of the density dependence of the coupling constants. 
The comparison between the two density dependencies is depicted in Fig.~\ref{fig:Compare_DD}. However, it remains difficult to draw any conclusion about a possible impact of the tensor term on the $b_i$ and $c_i$ values: except for the $\rho = 0$ value, the $\Gamma _\sigma (\rho)$ evolution is not much impacted by the parametrizations, although the asymptotic value at large $\rho$ is modified. In the case of $\Gamma _\omega (\rho)$, the function decreases slower, but also starts at lower value with DD-MEVT, compared to the DD-MEV case. The values at large $\rho$ are not much modified, with respect to DD-ME2-like interactions. It should be noted that $\Gamma _\omega (\rho)$ is almost constant with the density in the case of tensor interaction. The isovector sector is also slightly modified with an increase of the $\Gamma _\rho$ coupling constant, while the $a_{tv}$ parameter encoding the density dependence takes a smaller value w.r.t. DD-ME2-like interaction. However these modifications have a small impact on the behavior of $\Gamma _\rho (\rho)$.

\section{Application and discussion}

\subsection{Infinite nuclear matter}

The nuclear matter properties are studied by computing the different quantities previously introduced in Sec.II. 
The tensor term being a derivative quantity, it does not impact any of the infinite matter parameter. However, as previously shown, the presence of a tensor term modify the ending point of the fitting procedure and hence, shall modify nuclear matter properties.
Results are given in Tab.\ref{tab:nuc_matter}.

\begin{table}
    \begin{tabular}{  P{18mm} P{12mm} P{19mm} P{19mm} P{13mm}}
    \hline \hline
     & DD-ME2 & DD-MEV & DD-MEVT & Empirical datas\\
    \hline

    Dirac & 0.57 & \bf{0.58} & \bf{0.63} & 0.75 $\pm$ \\
     mass $M^\star/ M$ & & 0.58 (0.01) & 0.63 (0.01) & 0.1 \\

    Saturation & 0.152 & \bf{0.151} & \bf{0.152} & 0.155 $\pm$ \\
     density $\rho _{sat}$ (fm$^{-3}$) & & 0.152 (0.002) & 0.152 (0.002) & 0.005 \\

    Binding & -16.14 & \bf{-16.14}  & \bf{-16.14} & -15.8 $\pm$ \\
     energy per nucleon $E_0$ (MeV) & & -16.14 (0.02) & -16.16 (0.02) & 0.3\\

    Incomp-& 250.9 & \bf{238.5} &\bf{ 236.7} & 230 $\pm$ \\
     ressibility K$_\infty$ (MeV) & & 238.2 (3.8) & 235.1 (3.5) & 20 \\

    Skewness & -478 & \bf{-373} &\bf{ -287} & -700 $\pm$\\
     Q (MeV) & & -345 (36) & -257 (26) &  500 \cite{far97}\\
    \hline
    Symmetry & 32.3 & \bf{31.5} & \bf{31.4} & 32 $\pm$ \\
     energy J (MeV) & & 31.6 (0.3) & 31.6 (0.3) & 2\\

    Symmetry & 52.2 & \bf{44.3} & \bf{46.9} & 60 $\pm$ \\
     energy slope L (MeV) & & 44.0 (3.4) & 47.1 (3.7) & 15\\

    Symmetry & -87 & \bf{-57} & \bf{-72} & -100 $\pm$ \\
     energy incompressibility $K_{sym}$ (MeV) & & -53 (10) & -79 (6) & 200\\
    \hline \hline
    \end{tabular}
    \caption{Infinite nuclear matter parameters at saturation density, related to the isoscalar and isovector properties. Emprical datas are taken from \cite{mar18}. As in Tab.~\ref{tab:parameters}, bold numbers refer to the results obtained with the chosen parametrization, and below are written the mean values and uncertainties obtained over MCMC sampling.}
    \label{tab:nuc_matter}
\end{table}

Few differences appear between DD-ME2 and DD-MEV parametrizations: in the incompressibility $K _\infty$, which is decreased by more than 10 MeV and in the symmetry energy slope, which is also decreased by $\sim$1 MeV.
However, the corresponding values are similar between DD-MEV and DD-MEVT.
The main difference between DD-MEV and DD-MEVT lies in the Dirac mass value, defined as $M^\star = M - \Gamma _\sigma \sigma$. This result is explained by the smaller value of $\Gamma _\sigma$ between DD-MEV and DD-MEVT parametrizations, leading to an increased value of the Dirac mass.
This confirms that tensor term allows to get a larger value of Dirac mass, which is in better agreement with experiment\cite{typ20}.

\subsection{Binding-energy and radii}
\begin{figure}
    \centering
    \includegraphics[width=1\linewidth]{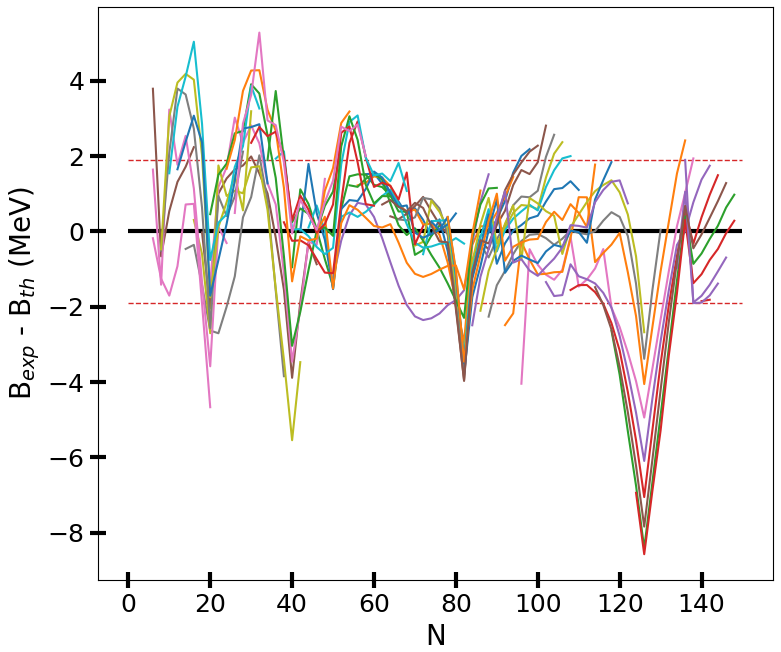}
    \caption{Evolution of $B_\text{exp} - B_\text{th}$ with neutron number. The dashed red lines represent the BE RMS.}
    \label{fig:dB_DDVT}
\end{figure}

The RMS values of the binding energies and radii are relevant quantities to test an interaction. These RMS values are computed over 828 nuclei with an axial code to test the different parametrizations and study the impact of the tensor terms. As previously discussed, the selected parametrization is chosen as the one minimizing the RMS values.
Results are shown in Fig.~\ref{fig:dB_DDVT} where $B_{exp} - B_{th}$ are plotted for isotopic chain up to Californium ($Z=98$), for the best fitted parametrization.

These results show the usual trend of isotopic drifts and peaks at $N=20,40,80,126$ with a  pronounced disagreement with experiment at $N=126$. 
The RMS values are summarized in Tab.\ref{tab:RMS}. It appears that the DD-ME2 interaction suffers from a large value of BE RMS mainly due to an important shift of 2 MeV between the theoretical predictions compare with experiment. 

\begin{table}
    \centering
    \begin{tabular}{ P{22mm} P{16mm} P{16mm} P{18mm}}
    \hline \hline
     & DD-ME2 & DD-MEV & DD-MEVT\\
    
    \hline

    $\overline{E_{exp} - E_{th}}$ & 2.0 & 0.1 & -0.02 \\

    $\sigma (E_{exp} - E_{th})$ & 2.88 & 2.10 & 1.89 \\

    $\sigma (R_{exp} - R_{th})$ & 0.26 & 0.25 & 0.25\\
    \hline \hline
    \end{tabular}
    \caption{Binding energy mean value and RMS (line 1 and 2) and radius RMS (line 3) for different parametrizations. Experimental values taken from Ref.\cite{hua16}}
    \label{tab:RMS}
\end{table}

\begin{figure*}
    \centering
    \includegraphics[width=1\linewidth]{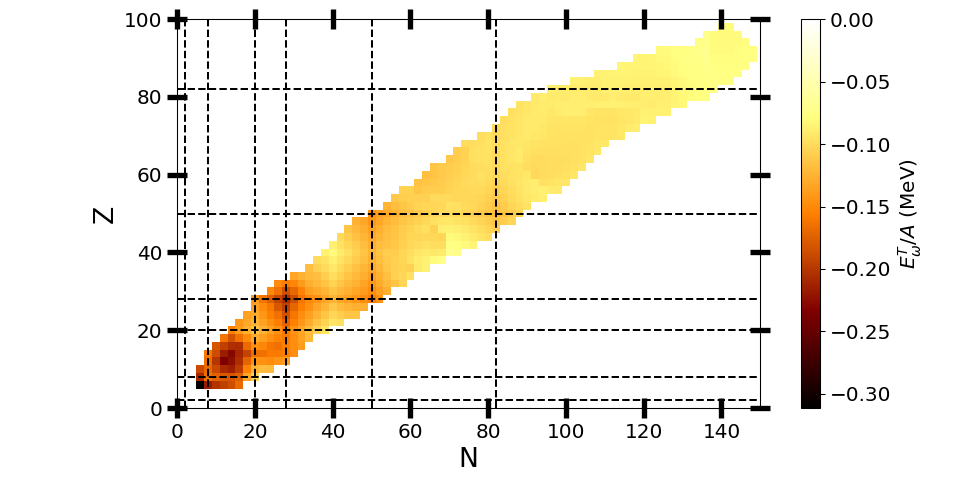}
    \caption{Evolution of the tensor energy per nucleon $E_\omega ^T / A$ with neutron and proton numbers.}
    \label{fig:E_w_T}
\end{figure*}

In order to study in more details the impact of the tensor term, Fig.~\ref{fig:E_w_T} shows the tensor energies as a function of neutron and proton number.
It is first interesting to notice that the tensor energy mainly contributes to light nuclei and becomes almost negligible for $N>100$.
A specific feature appears for the different neutron and proton numbers $N,Z=8,20$, where the tensor contribution gets smaller. This property can be understood in terms of spin saturation effect : due to the specific form of the tensor term, it should not contribute significantly for spin-saturated system. More precisely, shell closures at $N=8,20$ lead to a decrease of the tensor effect and thus to a reduced contribution to the total energy. 
However, the $N=28$ shell closure does not seem to impact the energy. This feature has already been found in other studies where the comparison between functional with and without tensor terms lead to similar results in these isotones \cite{mor10}.

A general feature of Fig.~\ref{fig:E_w_T} lies in the decrease of the contribution of the tensor term to the total energy with increasing mass number. This contribution becomes negligible for large mass nuclei where it does not exceed 0.05 MeV per nucleon. This property can be understood by considering that the tensor term is a surface term.

\subsection{Impact of deformation}

It can also be interesting to study the impact of tensor contribution to deformation. This study can be undertaken in several ways, but keeping control of the deformation remains the simplest solution. Hence, it is convenient to constrain the deformation of different nuclei and study the resulting potential energy surface (PES).

\begin{figure}
    \centering
    \includegraphics[width=1\linewidth]{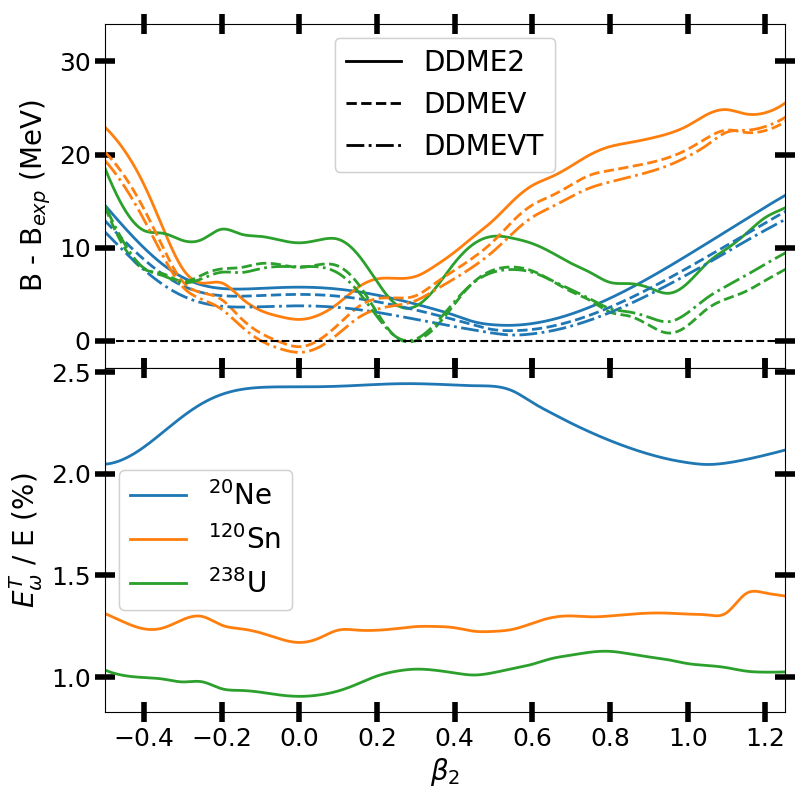}
    \caption{1D potential energy surface (upper panel) for $^{20}$Ne, $^{120}$Sn and $^{238}$U with DD-ME2, DD-MEV and DD-MEVT interactions. The ratio between the tensor energy contribution and the total energy is also plotted in the lower panel in the case of DD-MEVT interaction.}
    \label{fig:Tensor_PES}
\end{figure}

Fig.~\ref{fig:Tensor_PES} displays some PES for $^{20}$Ne, $^{120}$Sn and $^{238}$U nuclei which has been chosen to study the impact of tensor term on clusterized nucleus ($^{20}$Ne), and fission barriers ($^{238}$U).
First, the previously discussed result concerning the better description of measured masses by the DD-MEV and DD-MEVT functionals, compared to DD-ME2. 

In the case of $^{20}$Ne, the contribution of the tensor term is constant between $-0.2<\beta_2<0.6$ but undergoes a significant modification beyond these values, when the slope of the PES becomes steeper. It is interesting to note, however, that such a modification of the tensor part does not seem to have any influence on the total energy of the system.

In the case of $^{120}$Sn, the differences are negligible over the whole PES. However, the case of Uranium might be more relevant to study, as the $\beta _2$ barrier is slightly modified. First, its height is decreased by $\sim 1$ MeV, but its width is increased. 
Such a decrease in the height of the barrier, when adding the tensor term, has already been noticed  in a previous study \cite{tol18}. It remains rather difficult to draw conclusions from this feature, since a PES in $\beta _2$ and $\beta _3$ would be necessary to really compare the energies.

As already discussed in Fig.~\ref{fig:E_w_T}, the tensor contribution to the total energy is more important in light nuclei. Its evolution with deformation remains rather difficult to follow and depends on the detail of the spectrum.

\subsection{Gaps and spin-orbit splittings}

Gaps and spin-orbit splittings are often used to constrain and test the effect of tensor terms. In the present fitting procedure, we have mainly used gaps constraint since different studies previously mentioned that tensor terms have less (if not null) effect of SO splittings \cite{ren95,lon07}. Here, we propose to study the impact of tensor terms on both SO splittings and gaps.

Tab.\ref{tab:RMS_SO} displays RMS values for SO splittings, compared to experimental values for some spherical nuclei of Ref.\cite{lon07}. Associated standard deviation were computed using the first 1000 more converged MCMC calculations.
It should be noted that, already for DD-MEV interaction, which is equivalent to DD-ME2, the improvement is important ($\sim$20\%). The results are again improved by a factor of $\sim$15\% when the tensor term is added. It has been checked that the $\rho$ part of the tensor term does not contribute more than $0.02$ MeV for each SO splittings. Hence, the $\omega$ tensor term is the dominant tensor contribution.

The RMS gaps have been computed using 17 gaps (11 of them being in the constraints set of Tab.\ref{tab:constraints}) of Ref.\cite{mor10}. The RMS is reduced by $\sim$10\% between DD-ME2 and DD-MEV. This is due to the fact that more than $\sim$50\% of the gaps have been used in the constraint set. A second decrease of $\sim$20\% is observed when including the tensor term. Again, the presence of $\rho$ tensor term does not impact the gaps values.

This analysis shows that $\omega$ tensor term helps to decrease SO splittings and RMS gaps. However, it remains quite hard to analyse if this improvement directly comes from the tensor terms structure, or just from the addition of a new vector isoscalar d.o.f. 
In any case, it is clear that the $\rho$ tensor term does not impact the values of these RMS and thus do not contribute much at the Hartree level. However, it has been showed that the $\rho$ tensor terms have non-negligible contribution to these terms at the Hartree-Fock level \cite{jia05,wan13,wan18}.

\begin{table}
    \centering
    \begin{tabular}{  P{30mm} P{17mm} P{17mm} P{17mm}}
        \hline \hline
        RMS & DD-ME2 & DD-MEV & DD-MEVT \\
        \hline 
        SO splittings (MeV) & 0.70 & 0.59 (0.03) & 0.52 (0.01) \\

        Gaps (MeV) & 1.47 & 1.35 (0.01) & 1.12 (0.02)\\
        \hline \hline
    \end{tabular}
    \caption{RMS values for SO splittings and gaps for different interactions and nuclei of Ref.\cite{mor10}. Standard deviations are given in parenthesis. See the main text for details.}
    \label{tab:RMS_SO}
\end{table}

\subsection{Density}

Finally, the effect of tensor on the nucleus density is studied. It is expected that it may have an impact when the density rapidly changes, due to the derivative form of the tensor term. In particular, this might have important effects for cluster structures, bubble nuclei or neutron skin.

\begin{figure*}
    \centering
    \includegraphics[width=0.8\linewidth]{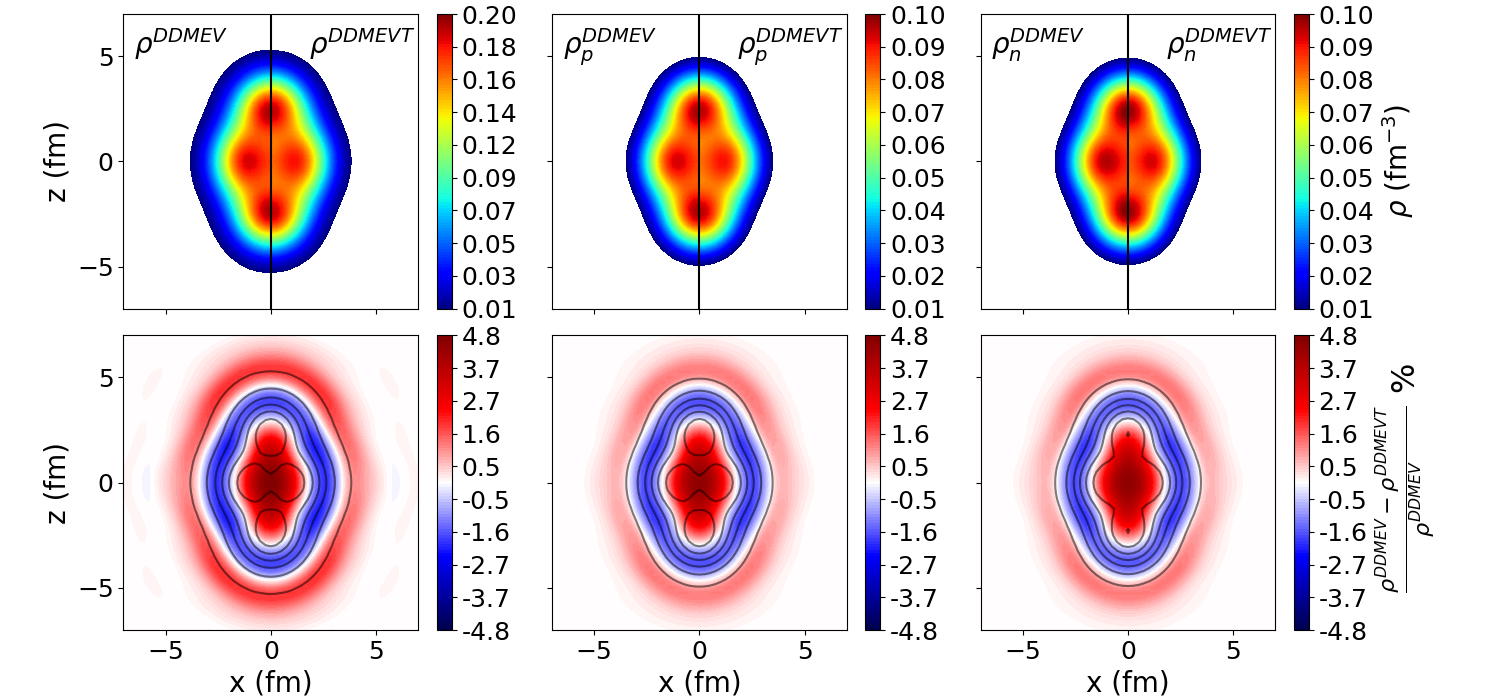}
    \caption{Total (left), proton (middle) and neutron (right) densities (upper panels) and density differences (lower panels) for $^{20}$Ne for DD-MEV and DD-MEVT interactions.}
    \label{fig:20Ne_dens}
\end{figure*}

Fig.~\ref{fig:20Ne_dens} shows the density and density difference of $^{20}$Ne between DD-MEV and DD-MEVT interactions. The cluster structures of $^{20}$Ne ground state makes this nucleus particularly interesting to study the impact of tensor term. In this specific case, a decrease of the density ($\sim$5\% of relative variation w.r.t. DD-MEV) is visible, mainly in the center, while an increase appears at the surface($\sim$2\% of relative variation).
The cluster structures seem less pronounced in the DD-MEVT case, and the largest density value drops from 0.191 fm$^{-3}$ to 0.187 fm$^{-3}$.

\begin{figure*}
    \centering
    \includegraphics[width=0.8\linewidth]{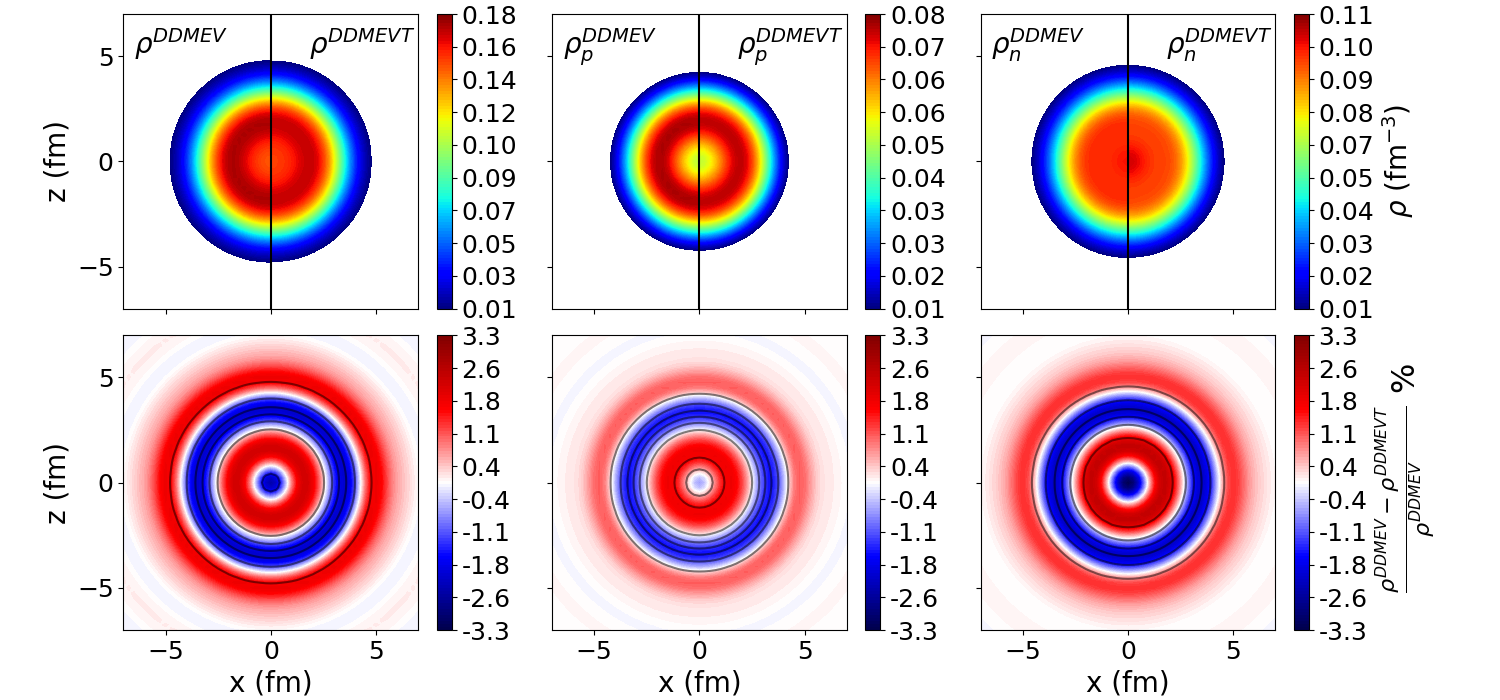}
    \caption{Same as Fig.~\ref{fig:20Ne_dens} but for $^{34}$Si nucleus.}
    \label{fig:34Si_dens}
\end{figure*}

Fig.~\ref{fig:34Si_dens} shows the same quantities for $^{34}$Si nucleus, which exhibit a bubble structure in the proton density. However, there are layers of small variations starting with an increase of the proton density in the center ($\sim$2\%) up to $\sim$1 fm, then a decrease up to 2 fm, etc. Nevertheless, the bubble in the proton density survives to the tensor term contribution.

\section{Conclusion}

The impact of adding a tensor term at the RHB level have been studied, by re-fitting a relativistic functional based on meson exchanges. 
For this purpose, data on single-particle gaps were added to the set of constraints. 
A first MCMC sampling method has been used in order to obtain information on local minima and compute reliable uncertainties. A second simplex minimization step was then implemented on top of the local minima found by the MCMC sampling.
It turned out that a parameterization minimizing the $\chi ^2$ related to set of constraints was not always favorable for testing the properties of the nuclei on a larger scale.

Results of this minimization show that the isoscalar tensor term ($\omega$ channel) is well constrained by our set, while the isovector part is not. This result can have two causes: the set of constraints does not allow to constrain this parameter, or the isovector tensor part of the interaction is negligible at the Hartree level (a result which has already been noticed in previous studies \cite{ruf88}).

At the finite nuclear matter level, adding a tensor term in the isoscalar part of the interaction, decreases the binding energy, gaps and spin-orbit RMS by $\sim 10-20\%$. The tensor term is much stronger in low-mass nuclei but can have non-negligible effects on potential energy surface of high-mass nuclei.

In the case of infinite matter, the main result is the increase of the Dirac mass, as expected. This increase is explained by the lowering of the coupling constant in the scalar-isoscalar channel, modifying the scalar potential used in the computation of the Dirac mass. This lowering of the $\sigma$ coupling constant is related to the increase of the spin-orbit splitting due the tensor contributions. This increase allows for a better agreement with experiment.

The Fock contribution is known to be important in order to take into account tensor effects in the covariant approach. In particular, the effect of the tensor $\rho$ meson is negligible at the Hartree level, while several studies, considering the Fock term, accurately constrained this parameter, with a non-negligible contribution from this part of the interaction.
Although, at the relativistic level, it remains numerically involved to perform RHFB calculations in deformed nuclei, it may be relevant, in a near future, to compare the results of the present study with deformed RHB calculations on selected nuclei.

\begin{acknowledgments}
The authors thank S.Typel for fruitful discussions.
\end{acknowledgments}

\newpage

\appendix

\section{Matrix elements in spherical case\label{app:sph}}

For systems possessing a rotational invariance, the spherical basis is used and defined by 
\begin{equation}
    x=r\sin\theta\cos\varphi,\ \ y=r\sin\theta\sin\varphi,\ \ z=r\cos\theta
\end{equation}
The fermion wavefunction are then written as  
\begin{equation}
    \psi_{i}\left(\boldsymbol{r}\right)=\frac{1}{r}\begin{pmatrix}f_{i}\left(r\right)\Omega_{l_i+1/2,l_i,m_i}\left(\theta,\varphi\right)\\
ig_{i}\left(r\right)\Omega_{l_i-1/2,l_i,m_i}\left(\theta,\varphi\right)
\end{pmatrix}
\end{equation}
where $ \Omega_{jlm}\left(\theta,\varphi\right)$ holds for the usual spin-spherical harmonics. It is then useful to define a new quantum number $\kappa=\mp\left(j+1/2\right)$ such that $\Omega_{\kappa m}=\Omega_{jlm}$.
The radial wavefunctions $f_i$ and $g_i$ are then expended in terms of radial functions of a spherical harmonic oscillator potential. For more details about the numerical implementation of the RHB equations, see \cite{nik14}. 

Dirac matrix elements $\mathcal B$ are related to the Hamiltonian of Eq.\eqref{eq:hd} through the non-diagonal part of the kernel $\mathfrak{h}_{kk} = \int d^3r \psi_k ^\dagger h_D \psi_k$. More precisely, we have
\begin{equation}
    \mathfrak{h}_{kk}^\mathcal{B} = \sum_{\alpha \tilde{\alpha}^\prime} f_\alpha ^k \mathcal{B}_{ \alpha     \tilde{\alpha}^\prime} g_{ \tilde{\alpha}^\prime}^k + \sum_{\alpha  \tilde{\alpha}^\prime} g_{\tilde{\alpha}}^k \mathcal{B}_{\tilde{\alpha} \alpha^\prime} f_{\alpha^\prime}^k
\end{equation}

In spherical symmetry, the Dirac matrix elements $\mathcal B$  can be written as 

\begin{equation}
    \mathcal{B}_{\alpha\tilde{\beta}}^{T,k} = -\int dr\rho\left(r\right)\delta_{\kappa_{\alpha}\kappa_{\tilde{\beta}}}\delta_{m_{\alpha}m_{\tilde{\beta}}}\frac{1}{r^{2}}\partial_{r}\left( \frac{f_{\phi}}{4M} r^{2}R_{\alpha}^{\star}R_{\tilde{\beta}}\right)
\end{equation}
where the $R _\alpha$ represents the spatial wavefunction and $\phi=\omega,\rho$. In our study the $f_{\phi}$ have been taken as constant and thus can be brought out of the derivative and integral.

Finally, the source term due to the tensor interaction will be given, assuming spherical symmetry, by 
\begin{equation}
    \vec{\nabla}\cdot\vec{T}_{k}=\frac{1}{r^{2}}\frac{\partial r^{2}T_{k}^{r}}{\partial r}
\end{equation}

\section{Matrix elements in axial case\label{app:ax}}

For system with axial symmetry, the axial basis is used and defined as 
\begin{equation}
x= r_{\perp}\text{cos}\left(\varphi\right),\ \
y= r_{\perp}\text{sin}\left(\varphi\right),\ \
z= z
\end{equation}
Since with axial symmetry the third component of the third component of the angular momentum is conserved and associated with the $\Omega$ quantum number, the fermion wavefunction are expanded as 
\begin{align}
\psi_{i}\left(\mathbf{r},s,t\right) & =\begin{pmatrix}f_{i}^{+}\left(r_{\perp},z\right)e^{i\left(\Omega_{i}-1/2\right)\varphi}\\
f_{i}^{-}\left(r_{\perp},z\right)e^{i\left(\Omega_{i}+1/2\right)\varphi}\\
ig_{i}^{+}\left(r_{\perp},z\right)e^{i\left(\Omega_{i}-1/2\right)\varphi}\\
ig_{i}^{-}\left(r_{\perp},z\right)e^{i\left(\Omega_{i}+1/2\right)\varphi}
\end{pmatrix}\chi_{t_{i}}\left(t\right)\nonumber \\
 & =\begin{pmatrix}F_{i}^{+}\left(r_{\perp},z,t\right)\\
F_{i}^{-}\left(r_{\perp},z,t\right)\\
iG_{i}^{+}\left(r_{\perp},z,t\right)\\
iG_{i}^{-}\left(r_{\perp},z,t\right)
\end{pmatrix}
\end{align}
Then, the fermion wavefunctions $f_i ^\pm$ and $g_i ^\pm$ are expanded in terms of eigenfunctions of a single-particle Hamiltonian for an axially deformed harmonic oscillator potential. For more details about the implementation of the RHB equations in axial symmetry, see \cite{nik14}.

The different matrix elements are computed in the following way: 

\begin{equation}
\mathcal{B}_{\alpha\tilde{\beta}}^{T}=\mathcal{B}_{\alpha\tilde{\beta}}^{T,+}+\mathcal{B}_{\alpha\tilde{\beta}}^{T,-}+\mathcal{B}_{\alpha\tilde{\beta}}^{T,3}
\end{equation}
where
\begin{eqnarray}
    \mathcal{B}_{\alpha\widetilde{\beta}}^{T,+}&=&\int d^{3}r\rho\delta_{ms_{\alpha},\downarrow}\delta_{ms_{\widetilde{\beta}},\uparrow}\partial_{+}\left(\frac{f_{\phi}}{4M}\phi_{\alpha}^{*}\phi_{\widetilde{\beta}}\right)\\
    \mathcal{B}_{\alpha\widetilde{\beta}}^{T,k,-}	&=&\int d^{3}r\rho\delta_{ms_{\alpha},\uparrow}\delta_{ms_{\widetilde{\beta}},\downarrow}\partial_{-}\left(\frac{f_{\phi}}{4M}\phi_{\alpha}^{*}\phi_{\widetilde{\beta}}\right)\\
    \mathcal{B}_{\alpha\widetilde{\beta}}^{T,k,3}	&=&\int d^{3}r\rho\delta_{ms_{\alpha},ms_{\widetilde{\beta}}}\partial_{3}\left(\frac{f_{\phi}}{4M}\phi_{\alpha}^{*}\phi_{\widetilde{\beta}}\right)
\end{eqnarray}

\section{Infinite nuclear matter properties \label{app:EoS}}
In order to constrain the parameter space and obtain information about infinite nuclear matter, usual thermodynamical quantities are considered. For completeness, we recall some of them in the following. More details about their derivations can be found in \cite{dut14,typ20}.

In infinite cold matter, the total energy is given by 
\begin{equation}
    \epsilon=\frac{E}{A}\rho = \frac{1}{2}C_{\sigma}\rho_{s}^{2}+\frac{1}{2}C_{\omega}\rho^{2}+\frac{1}{2}C_{\rho}\rho_{3}^{2}+\epsilon_{kin}^{p+n}
\end{equation}
and the kinetic energy is
\begin{equation}
    \epsilon_{kin}^{p,n}=\frac{1}{4}M^{\star}\rho_{s_{p,n}}+\frac{3}{4}E_{p,n}\rho_{p,n}
\end{equation}
where the $C_i$ are defined as $C_{i}=\frac{\Gamma_{i}^{2}}{m_{i}^{2}}$ and $E_{p,n}=\sqrt{M^{\star}+k_{F_{p,n}}^{2}}$ where $k_F$ refers to Fermi momentum.

The incompressibility and skewness coefficients are defined at saturation density by
\begin{equation}
    K=9\frac{dP}{d\rho} \Biggl|_{\rho _{sat}} ~~~ ; ~~~ Q=27\rho_{sat}^3 \frac{\partial ^3}{\partial \rho^3} \frac{E}{A} \Biggl|_{\rho _{sat}}
\end{equation}
Their full detailed forms can be found in Ref.\cite{dut14}.

Important quantities about the isovector part are the symmetry energy ($J$), its slope ($L$) and its second derivative ($K_{sym}$). 
The symmetry energy is given by 
\begin{equation}
    E_{sym}\left(\rho\right)=\frac{1}{2}\frac{\partial^{2}E\left(\rho,\delta\right)}{\partial\delta^{2}}\Biggl|_{\delta=0}
\end{equation}
where $\delta$ holds for the asymmetry parameter. Its expression reads 
\begin{equation}
    E_{sym}\left(\rho\right)=\frac{1}{6}\frac{k^{2}}{E}+\frac{1}{8}C_{\rho}\rho
\end{equation}
and $J=E_{sym}\left(\rho _{\text{sat}}\right)$. Its  slope and second derivative are simply defined as 
\begin{equation}
    L=3\rho_{sat}\frac{\partial E_{sym}}{\partial\rho} \Biggl|_{\rho _{sat}} ~~~;~~~ K_{sym}=9\rho_{sat}^{2}\frac{\partial^{2}E_{sym}}{\partial\rho^{2}} \Biggl|_{\rho _{sat}}
\end{equation}
Their full detailed forms can be found in Ref.\cite{dut14}.


\end{document}